\documentclass[prl,aps,twocolumn]{revtex4}
\usepackage{graphicx}
\usepackage{epstopdf}
\usepackage{color}
\usepackage{longtable}

\usepackage{txfonts}
\usepackage{natbib}

\begin{document}
\title{Magnetic Fields in the Large Scale Structure from Faraday Rotation measurements}

    \author{A. Neronov$^{1}$,
          D.Semikoz$^{2,3}$,
          M.Banafsheh$^{1}$
          }

   \affiliation{$^1$ISDC Data Centre for Astrophysics, Department of Astronomy, University of Geneva, Ch. d'Ecogia 16, 1290, Versoix, Switzerland 
              \email{Andrii.Neronov@unige.ch}\\
             $^2$APC, 10 rue Alice Domon et Leonie Duquet, F-75205 Paris Cedex 13, France \\
$^3$ Institute for Nuclear Research RAS, 60th October Anniversary prosp. 7a, Moscow, 117312, Russia
             \email{Dmitri Semikoz <dmitri.semikoz@apc.univ-paris7.fr>}
             }

\begin{abstract}
We search for magnetic fields outside galaxies and galaxy clusters by investigating redshift evolution of  Faraday rotation measures (RM) of extragalactic radio sources. Our analysis reveals a strong evidence for the redshift dependence of the mean of the absolute value of the RM. The evidence is further strengthened if  the Galactic contribution to the RM is subtracted. The hypothesis of the absence of the redshift evolution of residual RM is ruled out at $5\sigma$ level.  The observed redshift dependence of  RM is consistent with the possibility of the presence of nano-Gauss strength magnetic fields with correlation length shorter than 0.1~Mpc  in the weakly overdense elements of the Large Scale Structure traced by the Ly$\alpha$ clouds. 
\end{abstract}

\maketitle

{\it Introduction.}
Intergalactic magnetic field (IGMF) outside  galaxies and galaxy clusters might originate from the early Universe or be produced by baryonic outflows from galaxies at the latest stages of evolution of the Universe \cite{kronberg94,widrow02,durrer13}. Weakness of IGMF makes its detection challenging. Upper bounds on IGMF limiting the IGMF strength $B\lesssim 10^{-9}$~G are derived from  radio observations \cite{kronberg82,kronberg94,blasi99}. Lower bounds $B\gtrsim 10^{-17}$~G are obtained from $\gamma$-ray observations \cite{neronov10,tavecchio10,taylor11,dermer11,vovk12}.

The upper bound derived from radio observations is based on the non-detection of the  IGMF induced Faraday rotation of the polarization plane of radio emission from distant quasars.   The bound is derived from the upper limit on the Rotation Measure RM$=d\alpha/d(\lambda^2)$ ($\alpha$ the polarization angle, $\lambda$ is the wavelength), which is proportional to the integral $\mbox{RM}\propto \int n_e(\vec B\cdot d\vec l)$ of the product of the free electron density and the parallel component of magnetic field along the line of sight. 

IGMF is most probably frozen into the ionized  intergalactic medium (IGM) and scales with the plasma density as $B\sim n_e^{2/3}$ \cite{blasi99}. This means that the strongest contribution to the line-of-sight integral defining the RM is coming from the IGM regions with enhanced density and magnetic fields. Moderate overdensities with density contrast $\delta=m_pn_e/\rho_b\sim 1-100$ ($\rho_b$ is the average energy density of the baryons in the Universe and $m_p$ is the proton mass) are common and quasar lines of sight typically cross such overdensities (higher density "clouds"), as witnessed by the observations of the Ly$\alpha$ forests in quasar spectra \cite{rauch99}.  Amplification of magnetic field in these overdensities leads to the enhancement of RM~\cite{kronberg82}. Account of the amplification of magnetic field in the matter overdensities significantly changes the redshift evolution of the Faraday RM, speeding up the growth of the mean of the absolute value $\left<|RM|\right>$ in the redshift range $0<z<1$ and essentially flattening the evolution of $\left<|RM|\right>$ at $z>1$ \cite{blasi99} .  A correlation length dependent upper limit  2-3~nG on IGMF could be derived from the non-observation of such redshift evolution of RM \cite{blasi99}. In fact, the same type of measurements that constrain the IGMF strength also show the presence of non-negligible magnetic fields in the intervening clouds \cite{kronberg82}.

Evidence for correlation of the enhancement of the RM along the lines of sight passing through intervening structures was recently found in the RM data at 6 cm wavelength \cite{kronberg08,bernet08}, confirming previous observations with smaller data sets \cite{kronberg76,kronberg82}. The enhancements of the RM were found to correlate with the presence of Mg II absorbers along the lines of sight toward the observed quasars. The increased RM is then interpreted  as being the result of non-negligible magnetization of the intervening elements of the Large Scale Structure (LSS), which are possibly associated with galactic wind blown bubbles. This effect was not observed in the RM data at 21 cm wavelength \cite{bernet12}. This could be due to the presence of inhomogeneous Faraday Rotation screen, produced e.g. by the large scale magnetic field of the Milky Way or by magnetized clouds in the IGM  \cite{bernet12}. 

A large set of some $\sim 4\times 10^4$ RM measurements has been obtained from the NVSS  sky survey \cite{taylor09}. Some 10\% of those sources have redshift measurements \citep{hammond12}.   Larger data sample also allows a higher quality measurement of the Galactic contribution to the RM \cite{pshirkov11,opperman12,farrar12,farrar12a}, which needs to be subtracted from the overall RM measurement to get the Residual Rotation Measure (RRM) sensitive to the IGMF. 
Although the analysis of the Ref. \cite{hammond12} does not find an explicit redshift evolution of the RRM, it  reveals an anti-correlation of the RRM and the fractional polarization $p$. This effect could be interpreted as the result of de-polarization of the radio emission by the short correlation length fluctuations of the magnetic field and electron density in a set of magnetized intergalactic clouds between the sources and the Milky Way \citep{rees82,bernet12}.

In what follows we further explore the data set of Ref. \cite{hammond12} and study the redshift dependence of statistical characteristics of RM and RRM, namely, of the mean and median of the absolute values of these quantities. Such characteristics are more sensitive for the presence of IGMF than e.g. the variances $\left< \mbox{RM}^2\right>$ or $\left< \mbox{RRM}^2\right>$ \cite{blasi99}.  We find that the mean of the absolute value of  the RM, RRM increases with the redshift. Moreover, the redshift dependence of $\left<|\mbox{RM}|\right>$,  $\left<|\mbox{RRM}|\right>$ are consistent with those expected for IGMF frozen into the mildly overdense elements of the LSS traced by the Ly$\alpha$ clouds. 

{\it Redshift evolution of the RM.}
For our analysis we use a subset of the RM measurements extracted from the NVSS sky survey \cite{taylor09} with known redshifts  \cite{hammond12}.  This smaller RM catalog contains extragalactic sources at Galactic latitudes $|b|>20^\circ$ spanning wide redshift range $0<z<5.3$.

\begin{figure}
\includegraphics[width=\linewidth]{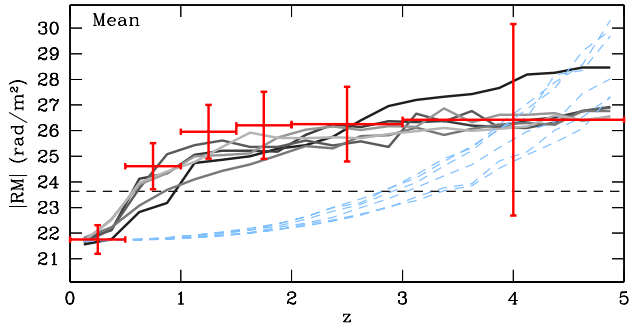}
\caption{Redshift dependence of the mean  of the absolute value of the RM in the full data set. Solid curves show the expected redshift dependences of the RM for the fields with different correlation lengths, $\lambda_B$, from  the Hubble radius (darkest curves)  down to 0.1~Mpc (lightest grey curves). Blue dashed lines show evolution in the models with magnetic field decreasing as $(1+z)^{2}$.}
\label{fig:all_allb}
\end{figure}

Fig. \ref{fig:all_allb} shows the redshift evolution of the mean of the absolute value of RM for the entire data set. In this figure one could clearly see the systematic increase of the $\left<|\mbox{RM}|\right>$ with the redshift up to $z\ge 1$. The best fit of the data with a constant, redshift independent RM (shown by the horizontal dashed line in the Figure) has the $\chi^2=25.04$ for 5 degrees of freedom. The chance probability for the redshift independent model to properly describe the data is $1.3\times 10^{-4}$. This means that no-redshift-dependence hypothesis is inconsistent with the data at $\simeq 3.7\sigma$ level. No significant redshift dependence is  present in the median of $|\mbox{RM}|$, which is readily explained by a somewhat  lower value of the median, which is difficult to measure on top of the foreground Galactic RM (see \cite{blasi99}).

\begin{figure}
\includegraphics[width=\linewidth]{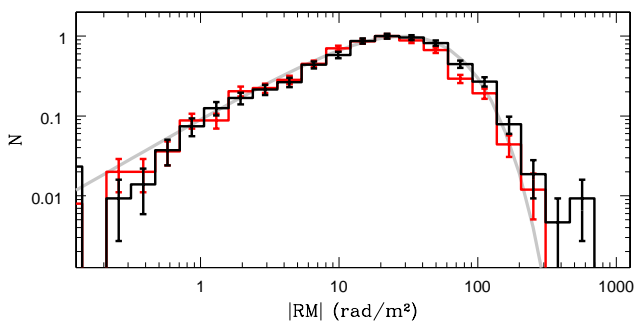}
\caption{RM  distributions at redshifts $0<z<0.5$ (red) and $z>1$ (black). Grey curve shows a fit with an exponential distribution.}
\label{fig:histogram}
\end{figure}

To clarify the nature of the increase of the mean $|\mbox{RM}|$ we plot in Fig. \ref{fig:histogram} the full distributions of $|\mbox{RM}|$ in two redshift bins: $0<z<0.5$ and $1<z<5.5$. The overall distributions are well fitted by an exponential distribution $dN/d(|RM|)\sim \exp(-|\mbox{RM}|/|\mbox{RM}|_0)$, where the width $|\mbox{RM}|_0$ is a parameter.  The higher values of the mean value of $|\mbox{RM}|$ at high redshifts is due to a slight difference in the $|\mbox{RM}|$ distributions, with an excess of high $|\mbox{RM}|$ values in the high redshift bin. 

At low Galactic latitudes, the RM of individual sources is dominated by the Galactic component. 
We have verified that the evidence for the redshift evolution of the $\left<|\mbox{RM}|\right>$ strengthens if one restricts the data selection to the sources at high Galactic latitudes. For example, considering only the data at $|b|>40^\circ$, one finds that the constant fit to the $\left<|\mbox{RM}|\right>(z)$ data gives a higher $\chi^2=28.4$ for 5 d.o.f. which corresponds to the chance probability $3\times 10^{-5}$ and detection of evolution at $\ge 4\sigma$ confidence level. This is expected if the evolution is due to the extragalactic contribution to the RM, which is more readily accessible in the lower Galactic RM regions at high Galactic latitudes. 

Solid lines in Fig. \ref{fig:all_allb} shows the result of model calculations of RM produced by the IGMF frozen in the LSS elements, with a constant, redshift independent, Galactic RM added in quadratures. To generate the model curves we have reproduced the Monte-Carlo simulations of IGMF induced RM from the Ref. \cite{blasi99}. In these simulations, the magnetic field strength is assumed to scale with the density of the medium  as $B\sim \rho^{2/3}$, as expected if the IGMF is frozen into the  collapsing / expanding regions of the intergalactic medium (IGM). To find the RM toward sources at different redshifts, one simulates the density profile of the IGM together with the associated IGMF along the lines of sight toward the sources. The density inhomogeneities occupy regions with the size equal to the redshift-dependent Jeans radius $R_J$. In each $R_J$ size region the overdensity is drawn from a log-normal distribution derived from the statistics of the Ly$\alpha$ clouds (see Ref. \cite{blasi99}). 

One could see that the redshift dependence of the RM produced by the IGMF frozen in the density inhomogeneities of the IGM provides a satisfactory description of the data. Different solid model curves in Fig. \ref{fig:all_allb} correspond to different assumed correlation lengths of IGMF. One could see that the shapes of the model curves for the $\left<|\mbox{RM}|\right>$ are almost correlation length independent, so that the observed evolution could be explained by the presence of magnetic fields with widely different correlation lengths.

For comparison, with dashed lines we show in Fig. \ref{fig:all_allb} model calculations for the case when the IGMF strength is assumed to be constant throughout the IGM (rather than scale with the density) and only evolve with the redshift as $B\sim B_0(1+z)^2$ as expected if the IGMF strength is only diluted by the expansion of the Universe.  One could see that such class of models clearly fails to reproduce the data, because the strongest contribution of IGMF to the RM in this case is expected for the sources at highest redshifts. 

\begin{figure}
\includegraphics[width=\linewidth]{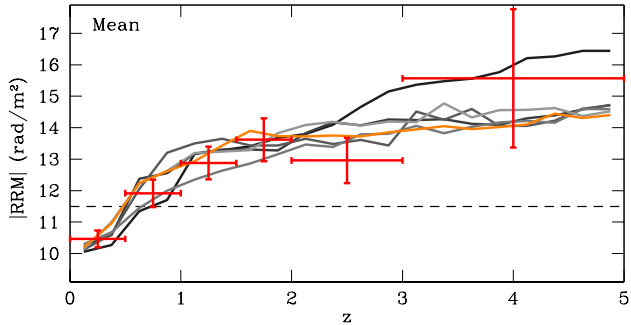}
\caption{Redshift dependence of the mean of the absolute value of the RRM in the full data set. Notations are the same as in Fig. 1.}
\label{fig:all_allb_rrm}
\end{figure}

The overall RM of extragalactic sources has a significant contribution from the Galactic magnetic field.  Subtraction of the Galactic contribution to the RM removes the background and makes the analysis of the signal produced by the IGMF easier. However, uncertainties in the knowledge of the Galactic magnetic field and of the distribution of the free electrons result in a direction-dependent uncertainty of the Galactic RM. The uncertainties introduced by the subtraction of the Galactic RM are difficult to characterize. A first estimate of the level of these uncertainties could be obtained e.g. by comparing different published estimates of the sky distribution of the Galactic RM, e.g. those obtained in the recent works \cite{opperman12,farrar12,farrar12a,pshirkov11} which agree in general large scale features of Galactic RM.  Subtracting the estimated Galactic RM \cite{opperman12} from the overall RM, one finds the distribution of RRM for the sources at different redshifts. Redshift evolution of the $\left<|\mbox{RRM}|\right>$ obtained in this way is shown in Fig. \ref{fig:all_allb_rrm}. 

Remarkably, subtraction of the Galactic RM does not remove the redshift dependence of the mean of the absolute value of RM. In fact, the inconsistency of the measurements with the hypothesis of redshift independent $|\mbox{RRM}|$ is increased. The $\chi^2$ of the fit of the data with constant is $\chi^2=39.9$ for 5 d.o.f., so that the redshift independent $\left<|\mbox{RRM}|\right>$ hypothesis is ruled out at $\ge 5\sigma$ level (chance probability for the constant model to describe the data is $p=1.6\times 10^{-7}$). 

Contrary to the overall $|\mbox{RM}|$, restriction of the data set to high Galactic latitude sources does not provide a stronger rejection of the no-evolution hypothesis. Taking only the data at 
$|b|\ge 40^\circ$
 one finds that the $\chi^2$ of the fit of $\left<|\mbox{RRM}|\right>(z)$ with a constant is $\chi^2=33.4$, i.e. somewhat lower than for the entire data set. This is naturally explained by the smaller number of measurements available at high Galactic latitudes.
 
{\it Investigation of possible selection biases.}
We have explored a number of possibilities for  the observed redshift dependence of the $\left<|\mbox{RM}|\right>$ and $\left<|\mbox{RRM}|\right>$ to arise in result of  selection bias in the dataset of Ref. \cite{hammond12}. 

The set of extragalactic sources under consideration has two main contributions from the  nearby radiogalaxies situated mostly at redshifts $z\le 0.5$ and QSOs mostly found at larger redshifts. Most of the redshift evolution of the RM takes place in the range of redshifts $0<z<1$. The dominant source class in different redshift bins changes in the range $0<z<1$ so that there is a possibility that the redshift dependence of the RM reflects the difference in the intrinsic properties of the sources. 

Such a possibility is, however, not consistent with the data. Indeed, measurement of the low value of  $\left<|\mbox{RM}|\right>$ and $\left<|\mbox{RRM}|\right>$ in the redshift bin $z<0.5$ limits the Galactic RM contribution to the RM of high redshift sources. If the excess RM at high redshifts is not due to the IGMF but is rather intrinsic to the QSO population, it is expected to strongly decrease with the redshift. The emission wavelength is related to the observed wavelength as  $\lambda_{em}=\lambda_{obs}/(1+z)$. The Faraday rotation angle at the emission wavelength is by a factor $(\lambda_{obs}/\lambda_{em})^2=(1+z)^{2}$ smaller for the sources at redshift $z$, so that the intrinsic source RM is expected to decrease as $(1+z)^2$. For a source at redshift $z\simeq 2$ the suppression is by more than an order of magnitude, so the intrinsic RM contribution is expected to become negligible at redshfits $z\ge 2$. This is not observed. 

A further verification of the independence of the result on the source class selection is found if only QSOs data are considered. Restriction of the data selection to QSO does not change the redshift evolution picture qualitatively. The main difference with the entire dataset is in the larger errorbars due to the lower number of sources.  Measurements of the $\left<|\mbox{RRM}|\right>$  in the individual redshift bins using only QSOs are consistent with  obtained using the entire data set. 

\begin{figure}
\includegraphics[width=\linewidth]{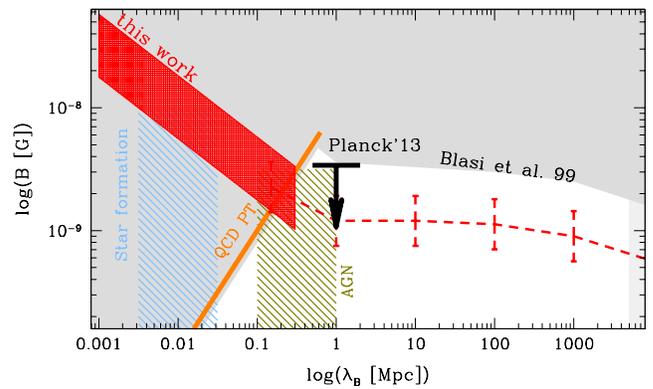}
\caption{Measurement of magnetic field  (red data points and red shading) compared to the constraints (grey shading) on IGMF strength and correlation length and model predictions (color) from Ref.  \citep{durrer13}.    Orange line marked "QCD PT" shows  range parameters for  magnetic field from QCD phase transition. Olive hatched region shows parameters of magnetic field from AGN radio lobes.  Black arrow is derived from the Planck data \citep{planck13}.  } 
\label{fig:constraints_Faraday}
\end{figure}

One more possibility for the redshift dependence to arise in result of selection bias is in the inhomogeneity of source distribution on the sky. For example, if high-redshift sources are preferably at low Galactic latitudes, where the $|\mbox{RM}|$ values are systematically higher, the redshift dependence of $\left<|\mbox{RM}|\right>$ might arise in result of inhomogeneous source distribution on the sky.  We have verified that this is not the case: the redshift evolution effect in the $\left<|\mbox{RM}|\right>$  is unchanged if sources at different Galactic latitudes are considered.

{\it Discussion.}
The observed redshift evolution of $\left<|\mbox{RM}|\right>$ confirms previous findings of the presence of magnetized medium between the sources and the Milky Way galaxy  \citep{kronberg76,kronberg08,hammond12}. It is satisfactory described by a model in which the IGMF is frozen in the LSS elements with moderate overdensities, so that $B$ scales as $\rho^{2/3}$ \cite{kronberg82,blasi99}.  The measured value of the magnetic field as function of the correlation length is shown in Fig. \ref{fig:constraints_Faraday} with red dashed line. 

A constraint on the size of the correlation length of magnetic field in the intervening overdense clouds could be derived from the knowledge of the angular/linear sizes of the sources and of the observation of depolarization of the radio signal in result of the Faraday rotation \citep{rees82,bernet12,hammond12}. We estimate the upper bound on the correlation length as $\lambda\lesssim 0.1-0.3$~Mpc, taking into account that the angular sizes of sources in the redshift range $z>1$ are $1''-10''$ (i.e. linear scales $\lesssim 0.1$~Mpc) and that the "Faraday screen" is situated at a comparable redshift $z\sim 1$. We plot the estimate of magnetic field $B=2\pm 1$~nG at the correlation length saturating this upper bound by the thick solid lines, while possible larger correlation length estimate by the dashed thin lines. Magnetic fields of the strength $B\gtrsim 10^{-9}$~G with correlation length much shorter than the upper bound $\lambda\lesssim 0.1$~Mpc would dissipate their energy on the time scales shorter than the Hubble time by driving plasma turbulence \citep{jedamzik04,durrer13}, if no turbulence is initially injected in the IGM. Note, however, that stronger fields with shorter correlation length could still be present in the gravitaitonally collapsing clouds in the IGM, because of the small-scale turbulence induced during the process of gravitational collapse \cite{Schleicher10}. 

Most of the observed RM signal  is accumulated when the lines of sight toward quasars pass through moderate overdensities (clouds) in the IGM \cite{kronberg82,blasi99}.  Magnetic fields in the regions with density contrast $\delta\le 1$ practically do not affect the signal, as noticed in the Ref. \cite{kronberg08}. We have verified this fact with the Monte-Carlo simulations described above. Setting the field strength to zero in the undersensities $\delta\le 1$ of the LSS (i.e. in the voids) does not change the redshift evolution of the RM. This means that the RM measurements are not sensitive to the IGMF in the voids of the LSS, such as the fields produced cosmologically \cite{kronberg94,widrow02,durrer13} or later at the onset of star formation activity \cite{miniati12}. The void magnetic fields could potentially be detectable with gamma-ray observations  \citep{plaga95,neronov07,neronov09}.

Figure \ref{fig:constraints_Faraday} shows predicted values of primordial magnetic fields from QCD (orange line)  phase transition  (the grey shadowed region corresponds to parameters  inconsistent with observational constraints, see review \cite{durrer13} for details). The estimated field strength is consistent with relic fields with the present-day correlation length $\lambda\simeq 0.1$~Mpc. This relatively large correlation length (for the given reference value of magnetic field strength) is the result of turbulence development which has removed the magnetic field power at smaller distance scales \cite{jedamzik04}. 

The magnetic fields in the walls / filaments / nodes of the LSS could be either a pre-existing magnetic field amplified in the course of structure formation, as suggested by the cosmological magnetogenesis scenaria (see \cite{durrer13} and references therein) or it could be the field spread across the LSS by galactic winds driven by the active galactic nuclei (AGN) and star formation activity \citep{bertone06,bernet08,kronberg08}. These two possibilities could be distinguished using the CMB data. Indeed, nG-strength IGMF of cosmological origin is expected to produce measurable effects on the CMB anisotropy and/or polarization angular power spectrum. 
CMB data provide tight constraints on normalization of the magnetic field which, in general, depend on the assumed slope of the magnetic field power spectrum $n$. The latest upper limit on $B$ marginalized over $n$ is reported by the Planck collaboration and is shown in Fig. \ref{fig:constraints_Faraday} as a downward pointing arrow at the reference distance scale $\lambda=1$~Mpc occasionally used in the data analysis procedure \cite{planck13}.
 This upper limit is just about our estimate of the field strength estimated above from the Faraday rotation data if one adopts the assumption of the cosmologically produced field. A viable mechanism of generation of such field would be production at the  QCD phase transition with subsequent amplification by turbulence at the moment of generation \cite{sigl94} or later during cosmological evolution \cite{sigl13}. The closeness of the CMB upper limits with the estimate from the Faraday rotation implies that the hypothesis of the cosmological origin of the observed nG field could possibly be tested  with a higher sensitivity analysis of the Planck data,  \cite{planck13}, specifically for the causally produced fields with $n=2$ \cite{durrer03} and taking into account non-linear effects \citep{shaw12}. 

{\it Acknowledgements.} We would like to thank P.~Kronberg and F.Miniati for useful comments on the text. The work of AN is supported by Swiss NSF grant PP00P2\_123426.

\end{document}